\documentclass{PoS}
\usepackage{cite}

\title{NLO mass effects in ${\bf b\bar{b}b\bar{b}}$ production at the LHC}

\ShortTitle{NLO mass effects in $b\bar{b}b\bar{b}$ production at the LHC}

\author{\speaker{Ma\l{}gorzata Worek} \thanks{Preprint number: TTK-13-23}\\
        Institute for Theoretical Particle Physics and Cosmology\\
        RWTH Aachen University\\ D-52056 Aachen, Germany \\
        E-mail: \email{worek@physik.rwth-aachen.de}}

\abstract{  The calculation of NLO QCD corrections to the production
  cross section of two bottom pairs 
  at the LHC  is briefly summarized.  We describe the $pp\to
  b\bar{b}b\bar{b}+X$ process both in the 4-flavor and  5-flavor scheme
  and investigate  the effect of the  finite bottom quark mass.  The
  results for the total cross section and a few differential
  distributions are presented. They have been obtained with the
  Nagy-Soper subtraction formalism for real radiation at NLO.  A
  comparison with results based on the traditional Catani-Seymour
  subtraction is also included. }

\FullConference{11th International Symposium on Radiative Corrections 
               (Applications of Quantum Field Theory to Phenomenology) 
               (RADCOR 2013),\\
		22-27 September 2013\\
		Lumley Castle Hotel, Durham, UK }

\begin{document}


\section{Introduction}


After the discovery of the Higgs boson at the Large Hadron Collider
(LHC)  by the ATLAS \cite{Aad:2012tfa} and CMS
\cite{Chatrchyan:2012ufa} experiments, a precise determination of
its properties is of extreme importance.  The determination of  the
Higgs boson couplings to fermions and gauge bosons as well as the
reconstruction of the Higgs potential are among the measurements that
are carried out.  The $b\bar{b}b\bar{b}$ final state can play
an important role in Higgs boson  studies at the LHC. For
instance, the
reconstruction of the Higgs potential requires the measurement of the
trilinear Higgs couplings that can be  performed in the $pp \to HH \to
b\bar{b}b\bar{b}$ channel
\cite{Djouadi:1999rca,Dolan:2012rv,Baglio:2012np}.  Moreover  the $pp
\to b\bar{b}H \to b\bar{b}b\bar{b}$ production mode where the  Higgs
boson is  radiated off a bottom quark could  be used to measure the bottom
quark Yukawa coupling \cite{Dittmaier:2003ej,Dawson:2003kb}.  This
final state is also of great significance in probing New Physics
scenarios, where for example a search for a model-independent s-channel TeV
resonance, that decays into a pair of Standard Model (SM) resonances, {\it
  e.g.} $Z$ or $H$, which subsequently decay into  four bottom
quarks \cite{Gouzevitch:2013qca}, could be performed. 
Accurate knowledge of the SM  background would play a crucial
role in devising strategies to look for physics beyond the SM. 

In QCD, the $pp \to b\bar{b} b\bar{b}$ process can be described either in
the four flavour scheme  (4FS) or in the five flavour scheme (5FS). In
the former case bottom quarks appear only in the final state and are
massive. They do not enter in the computation of the running of
$\alpha_s$ and the evolution of the PDFs. Finite  $m_b$ effects enter
via power corrections of the type  $\left[(m_b^2/Q^2)\right]^n$ and
logarithms of the type $\left[\log^{n}(m_b^2/Q^2)\right]$ where $Q$ stands for
the hard scale of the process. At the LHC, typically $(m_b/Q) \ll 1$
and power corrections are suppressed, while logarithms, both of
initial and final state nature, could be large. However,  for inclusive
observables such as b-jets, logarithms can only originate from nearly
collinear initial-state $g\to b\bar{b}$ splitting. These large
logarithms could in principle spoil the convergence of  fixed order
calculations and  a resummation could be required.  But 
up to NLO accuracy
those potentially large logarithms, $\log(m_b/Q)$, are  replaced by
$\log(p^{\rm min}_{\rm T, b}/Q)$ with  $m_b \ll p^{\rm min}_{\rm T,b}
\lesssim Q$ and are less significant  numerically. On the other hand,
in the 5FS towers of $\log^n(m_b^2/Q^2)$ can be explicitly resummed
into the bottom quark PDFs. For consistency with the factorization theorem, one
should set $m_b$ to zero in the calculation of the matrix
element. Therefore the number of active flavors is $N_F$ = 5 and
bottom quarks enter in the computation of the running of $\alpha_s$
and  evolution of the PDFs. To all orders in perturbation theory those
two schemes are identical  in describing logarithmic  effects.
However, the way of ordering in the perturbative expansion is
different and at any finite order the results might be different,  see
{\it e.g} \cite{Harlander:2011aa,Maltoni:2012pa,Frederix:2012dh}. 

NLO calculations for the $pp \to b\bar{b}b\bar{b} +X$ production   in
the 5FS with massless bottom quarks have   been first performed  by the
\textsc{Golem} collaboration  \cite{Binoth:2009rv,Greiner:2011mp}. We
have calculated this process  using both schemes, 4FS and 5FS, which
gave us an opportunity to study the impact of dominant mass
contributions \cite{Bevilacqua:2013taa}. We have also used this
process as a testing ground  to cross-check our  implementation of the newly
implemented Nagy-Soper subtraction scheme for both massive and
massless cases \cite{Bevilacqua:2013iha}.

In the following we briefly summarize the calculation of the NLO
corrections to the $pp \to b\bar{b}b\bar{b} +X$ process  at the LHC 
in the  4FS and the 5FS. In addition,  a comparison with results
calculated using  the traditional Catani-Seymour subtraction scheme will
also be presented.


\section{HELAC-NLO Framework}


Calculations are performed with the help of \textsc{Helac-NLO}
\cite{Bevilacqua:2011xh}, which is  based  on the tree level
\textsc{Helac-Phegas} framework
\cite{Kanaki:2000ey,Papadopoulos:2000tt,Cafarella:2007pc}.
The package consists of \textsc{Helac-1loop}  \cite{vanHameren:2009dr}
for the computation  of the one loop amplitudes, \textsc{CutTools}
\cite{Ossola:2007ax}, which implements the OPP reduction method
to decompose loop integrals into scalar integrals
\cite{Ossola:2006us,Ossola:2008xq,Mastrolia:2008jb,Draggiotis:2009yb}, 
and \textsc{OneLoop}
\cite{vanHameren:2010cp} for the evaluation of the scalar
integrals. The singularities for soft and collinear parton emission
are treated using two subtraction schemes as implemented in
\textsc{Helac-Dipoles} \cite{Czakon:2009ss}, namely Catani-Seymour 
\cite{Catani:1996vz,Catani:2002hc} and Nagy-Soper \cite{Bevilacqua:2013iha}
subtraction schemes. The idea for the latter subtraction scheme has
been first introduced by Nagy and Soper 
in the formulation of an improved parton shower \cite{Nagy:2007ty} and later on
exploited in a series of papers 
\cite{Chung:2010fx,Chung:2012rq,Robens:2013wga}.
The phase space integration is performed with the help of the Monte
Carlo generator  \textsc{Kaleu} \cite{vanHameren:2010gg}, including
\textsc{Parni} \cite{vanHameren:2007pt}  for importance sampling. The  
\textsc{Helac-NLO} package has already been widely used in the  computation 
of NLO QCD
corrections to several processes  at the LHC and the Tevatron
\cite{Bevilacqua:2009zn,Bevilacqua:2010ve,Bevilacqua:2010qb,Bevilacqua:2011aa,
Worek:2011rd,Bevilacqua:2012em}.


\section{Numerical Results for the LHC}


%
In the following we present predictions for the $b\bar{b}b\bar{b}+X$
process  at the LHC with $\sqrt{s}=14$ TeV.  All final-state
partons with pseudorapidity $|\eta| < 5$  are recombined into jets
with a resolution parameter $R = 0.4$ via the IR-safe anti-$k_T$ jet 
algorithm \cite{Cacciari:2008gp}. The four b-jets are required to have
\begin{equation}
p_T (b) > 30 ~{\rm GeV},	
~~~~~~~~~~~|y(b)| < 2.5,	
~~~~~~~~~~~\Delta R(b,b) > 0.4 \,,
\end{equation}
where $p_T (b)$ and  $y(b)$ are the transverse momentum and rapidity of
the  b-jet, whereas  $\Delta R(b,b)$ is the separation in the plane of
rapidity  and azimuthal angle between $b\bar{b}$ pairs. The five and 
four flavor 
MSTW2008 sets of parton  distribution functions (PDFs) 
are employed \cite{Martin:2009iq,Martin:2010db}.   In
particular, we take  MSTW2008LO PDFs with 1-loop running $\alpha_s$ in
LO and MSTW2008NLO PDFs  with 2-loop running $\alpha_s$ in NLO.  The
renormalization and factorization scales are set to a common value 
\begin{equation}
\mu_R=\mu_F=H_T= \sum m_{T}(b), ~~~~~~~~~~~~~~
m_{T}(b)=\sqrt{m^2_b+p^2_{T}(b)}.  
\end{equation}
For the four flavour scheme we define the bottom quark mass in the
on-shell scheme and use  $m_b = 4.75$\,GeV.
%
\begin{figure}
\begin{center}
\includegraphics[width=0.7\textwidth]{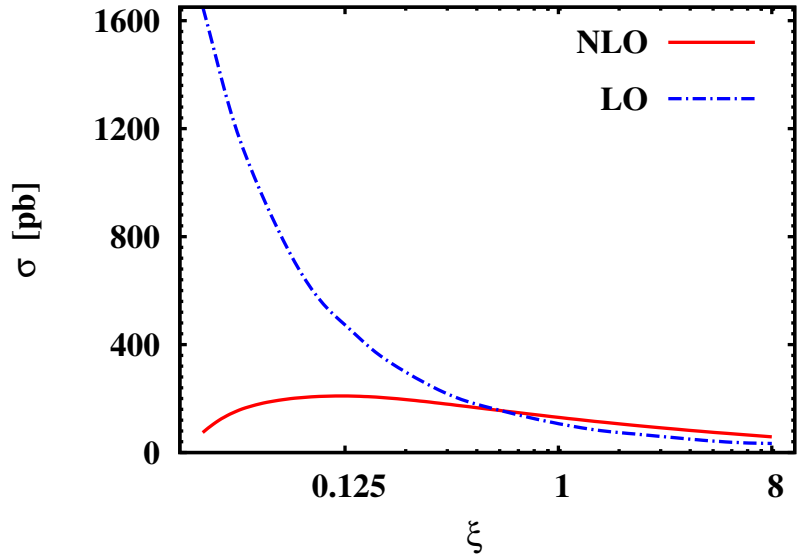}
\end{center}
\caption{\it \label{fig:scale-dependence} Scale dependence  of the 5FS
  LO and NLO cross sections for  $pp\rightarrow b\bar{b} b\bar{b} ~+
  X$ at the LHC  with $\sqrt{s}$ = 14 TeV. The renormalisation and
  factorisation  scales are set to a common value $\mu_R=\mu_F= \xi \mu_0$, 
  where $\mu_0=H_{T}$. }
  \end{figure}
%

\subsection{Comparison between 5FS and 4FS}

%
The cross section predictions in 5FS and 4FS are collected in Table
\ref{tab:1}. At the central value of the scale both cross sections
receive moderate  NLO correction of the order of $40\%$.  The scale
dependence is indicated by the upper and lower indices. The upper
(lower) index represents the change when the scale is shifted towards
$\mu = H_T/2 ~(\mu = 2 H_T)$. Rescaling the common scale from the
default value up and down by a factor 2 changes both cross sections at
LO by about $60\%$. Through the inclusion of NLO QCD corrections
scale uncertainties are reduced down to about $30\%$.  In Figure
\ref{fig:scale-dependence}  a graphical presentation of the scale
dependence is given, both at the LO and NLO. We observe a dramatic
reduction of the scale uncertainty while going from LO to NLO.
%
\begin{table}[h!]
\renewcommand{\arraystretch}{1.5}
\begin{center}
  \begin{tabular}{|c|c|c|c|}
\hline
$pp\to b\bar{b}b\bar{b}+X$& $\sigma_{\rm LO}$\,[pb] 
& $\sigma_{\rm NLO}$\,[pb] 
& $K = \sigma_{\rm NLO}/\sigma_{\rm LO}$ \\ \hline 
MSTW2008LO/NLO (5FS) & $99.9^{+58.7\,(59\%)}_{-34.9\,(35\%)}$ 
& $136.7^{+38.8(28\%)}_{-30.9\,(23\%)}$ 
& 1.37\\
MSTW2008LO/NLO (4FS) & $84.5^{+49.7(59\%)}_{-29.6(35\%)} $ & 
$118.3^{+33.3(28\%)}_{-29.0(24\%)}$ & 1.40\\
\hline
  \end{tabular}
\end{center}
  \caption{\it \label{tab:1} 5FS and 4FS LO and NLO cross sections for
    $pp\rightarrow b\bar{b} b\bar{b} ~+ X$ at the LHC with $\sqrt{s} =
    14  ~TeV$.   The
    theoretical uncertainties  and the K-factor are also given.}
     \end{table}
%
\begin{figure}
\begin{center}
\includegraphics[width=0.49\textwidth]{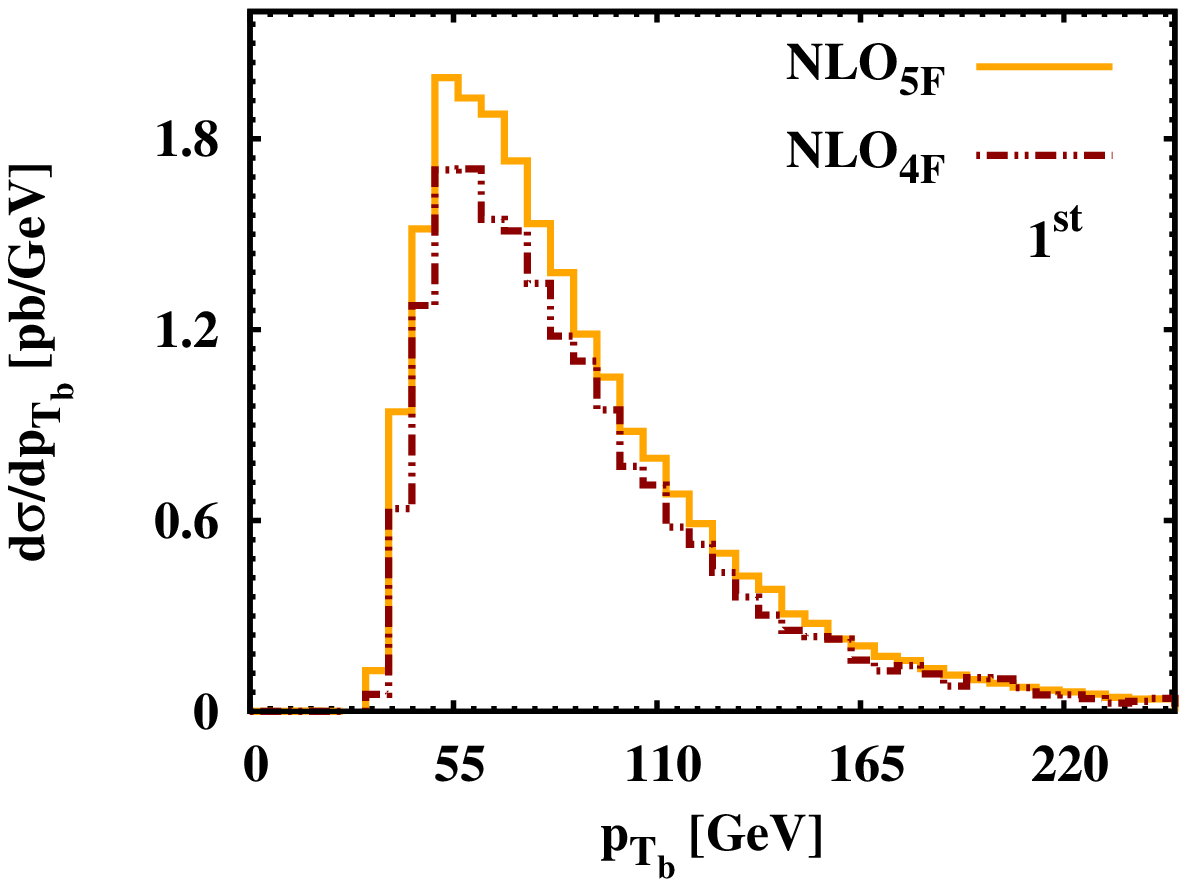}
\includegraphics[width=0.49\textwidth]{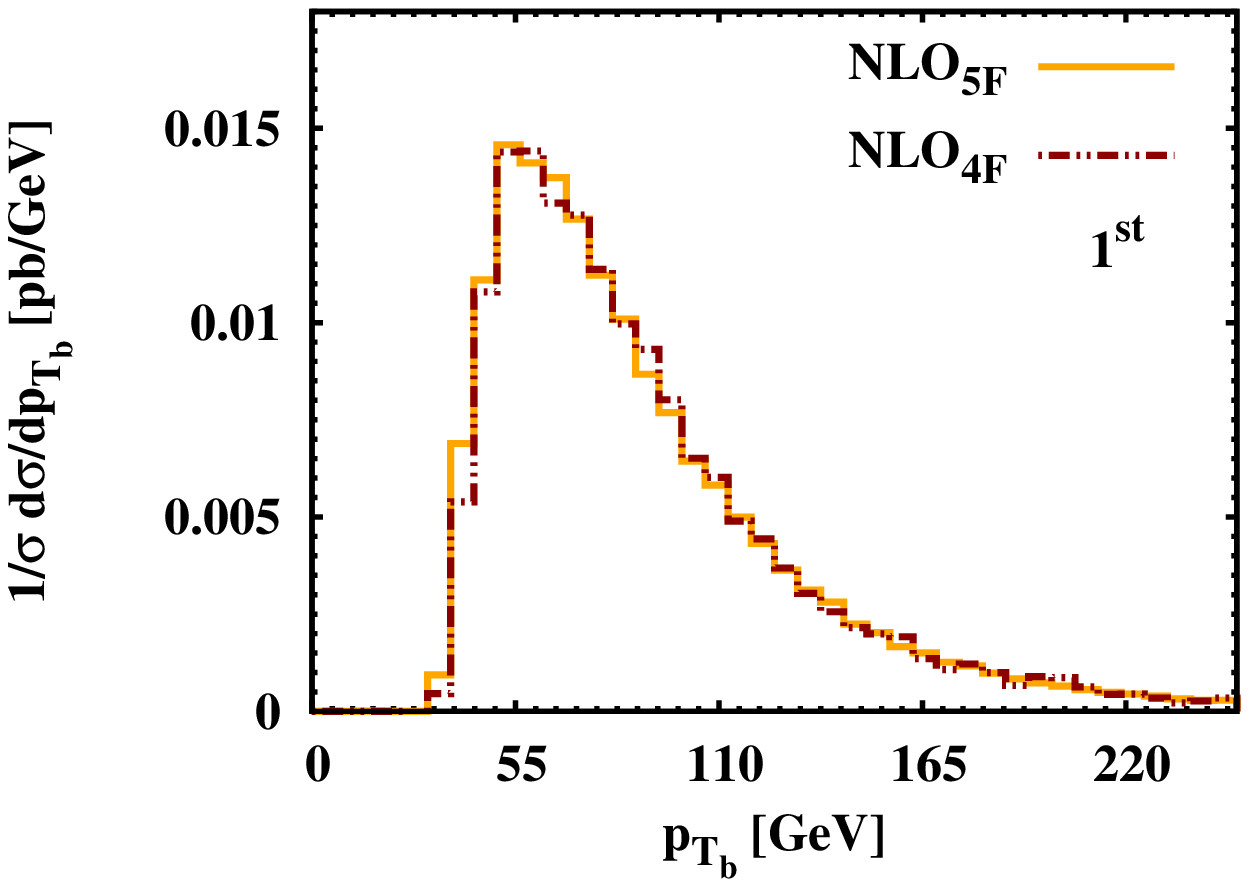}
\end{center}
\caption{\it \label{fig:1} Differential NLO cross section for
  $pp\rightarrow b\bar{b} b\bar{b} ~+ X$ at the LHC with $\sqrt{s}$ = 14
  TeV in the 4FS and 5FS  as a function of the transverse momentum of
  the hardest bottom jet. Also shown are the  normalised distributions
  at NLO.} 
  \end{figure}
%

Comparing 4FS with 5FS results, we observe that the
bottom  mass effects decrease the NLO cross section prediction by 
$16\%$. The difference between the massless  and the
massive calculations  has two origins.  First, we have a genuine
bottom mass effect of the
order of $10\%$ that  depends strongly on the transverse momentum cut 
and decreases to $1\%$ for $p_T (b)$ higher than $100$ GeV.
The remaining $\sim 6\%$ variation is due to
an interplay of two factors, different pdf sets and
different corresponding $\alpha_s$. The 4FS pdf set does not
comprise  $g\to b\bar{b}$ splitting therefore the corresponding
gluon flux is much
larger than for the 5FS pdf set. On the other hand, the four flavor 
$\alpha_s$ is  smaller than the corresponding value
calculated with five active flavors.  For the $pp\to b\bar{b}b\bar{b}
+X$ process the difference in $\alpha_s$  dominates, which accounts for 
a further reduction of the 4FS cross section  prediction.

An important input for the experimental analyses and the
interpretation of the experimental data are accurate predictions of
differential distributions. In Figure\,\ref{fig:1} the differential
distribution in the transverse momentum of the hardest bottom jet, as
calculated in the 5FS with massless bottom quarks and in the 4FS with
$m_b = 4.75$\,GeV is presented.  Both, the absolute prediction at NLO,
and the predictions  normalized to the corresponding 5FS and 4FS NLO
inclusive cross sections are shown. The latter plots make it clear
that the difference in the shape of the distributions in the 5FS and the
4FS is not significant.


\subsection{Comparison between Catani-Seymour and Nagy-Soper subtraction scheme}

%
The calculations have been performed with two different subtraction
schemes: the standard Catani-Seymour dipole subtraction, and a new
scheme based on the splitting functions and momentum mappings of an
improved parton shower by Nagy and Soper.  The comparison between the
two schemes for the inclusive 5FS and 4FS cross sections is presented
in Table \ref{tab:CS-NS}. Cross sections obtained using both
subtraction schemes are in agreement. They provide a validation of
our implementation of the new subtraction scheme into
\textsc{Helac-Dipoles} for the case of massive and massless fermions
and allow for  a non-trivial internal cross check of the calculation. 
\begin{table}[h!]
\renewcommand{\arraystretch}{1.5}
\begin{center}
  \begin{tabular}{|c|c|c|}
\hline
$pp\to b\bar{b}b\bar{b}+X$  
& $\sigma_{\rm NLO}^{\rm CS }$ [pb] & 
$\sigma_{\rm NLO}^{\rm NS}$ [pb] \\ \hline
MSTW2008NLO (5FS) & $136.7 \pm 0.3$ & $137.6 \pm 0.5$\\
MSTW2008NLO  (4FS) & $118.3 \pm 0.5$ 
& $118.0 \pm$ 0.5 \\
\hline
  \end{tabular}
\end{center}
  \caption{\it \label{tab:CS-NS} 5FS and 4FS NLO cross sections for
    $pp\rightarrow b\bar{b} b\bar{b} ~+ X$ at the LHC  with $\sqrt{s}$
    = 14 TeV. Results are shown for two different subtraction schemes,
    the Catani-Seymour (CS) dipole subtraction and the new Nagy-Soper
    (NS) scheme. The numerical error from the Monte Carlo
    integration is also included. }
 \end{table}
%
\begin{figure}
\begin{center}
\includegraphics[width=0.49\textwidth]{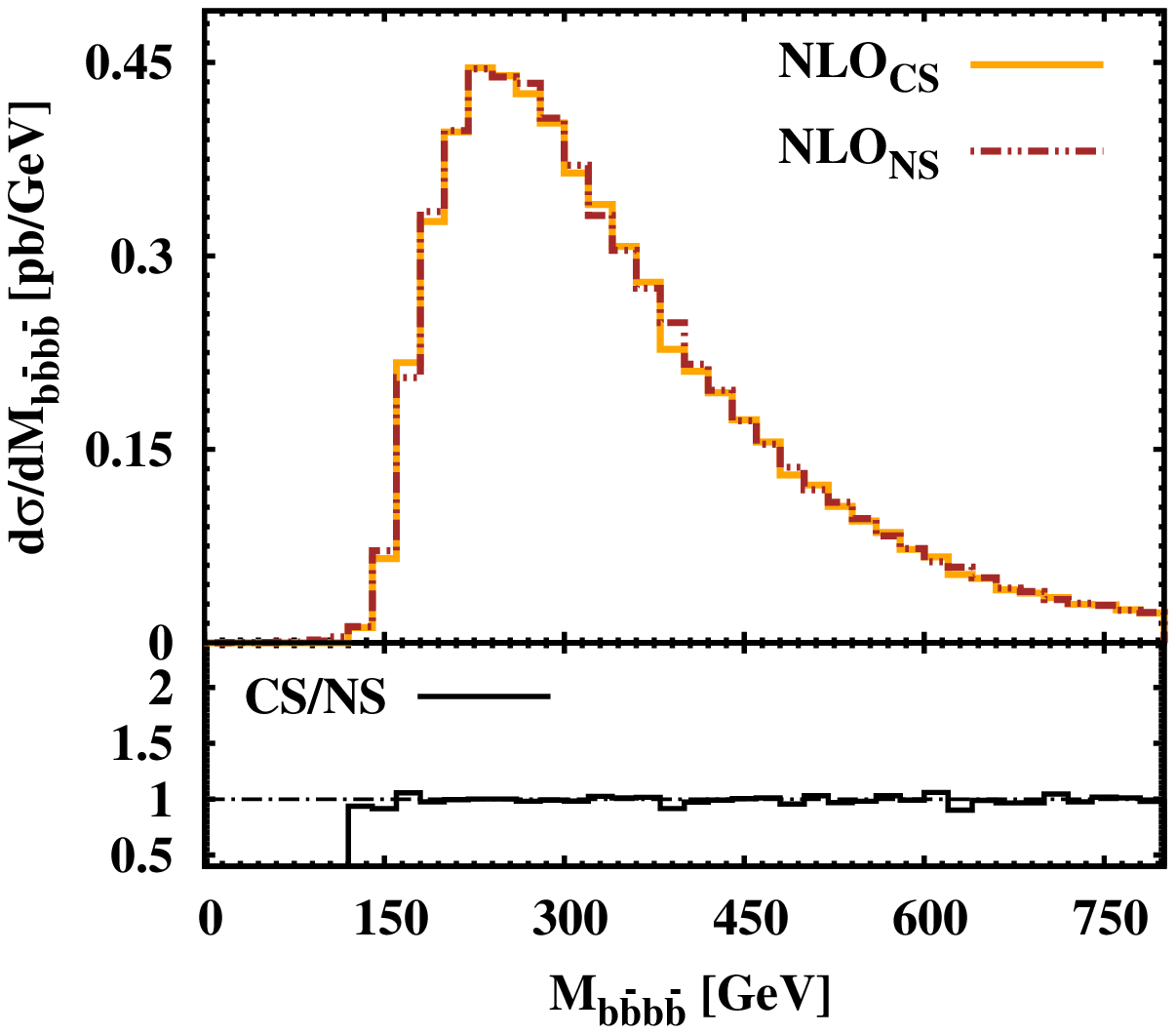}
\includegraphics[width=0.49\textwidth]{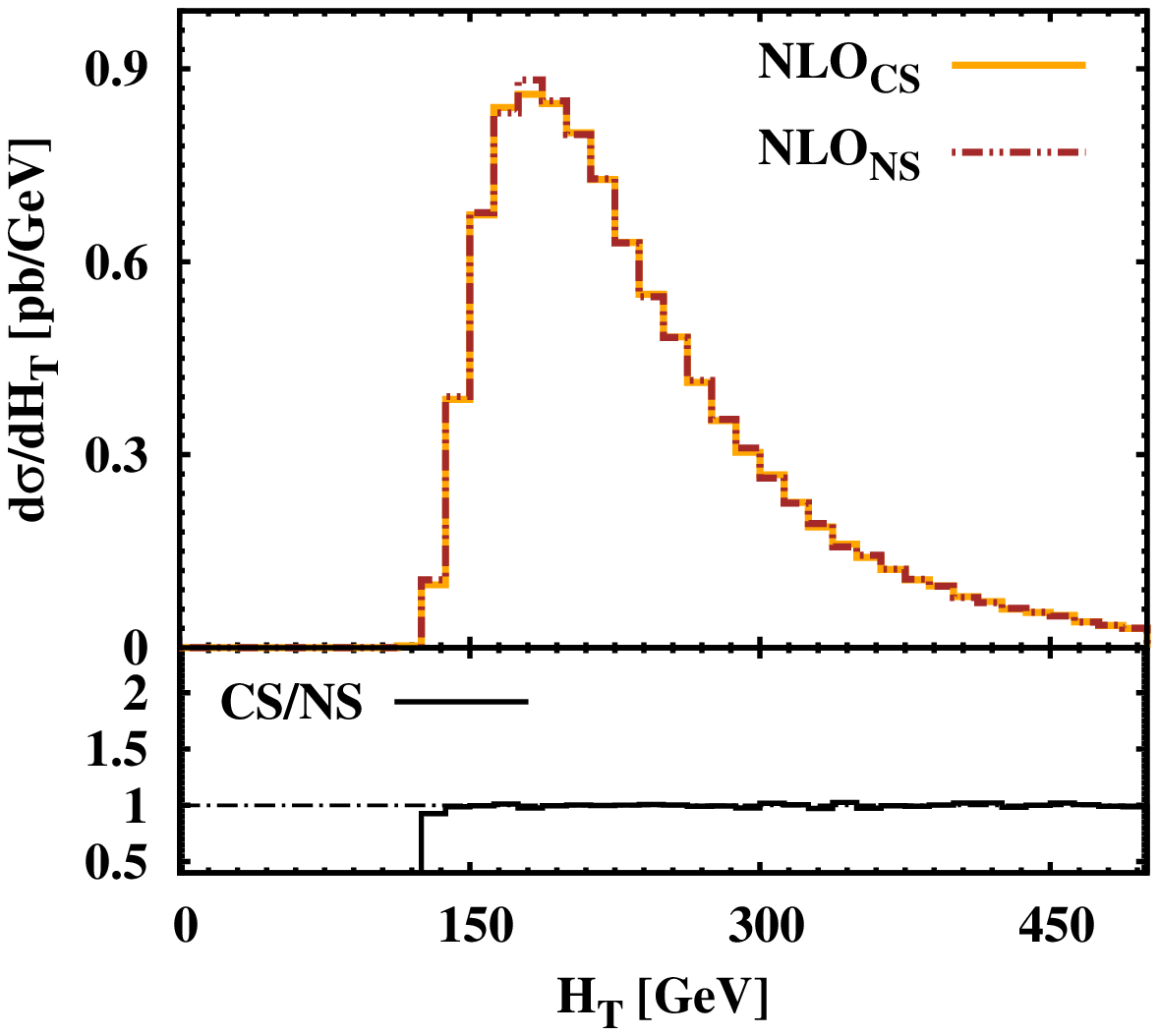}
\end{center}
\caption{\it \label{fig:CS-NS}   Differential cross section for
  $pp\rightarrow b\bar{b} b\bar{b} ~+ X$ at the LHC  with $\sqrt{s}$ =
  14 TeV in the 5FS as a function of  the $b\bar{b}b\bar{b}$ invariant
  mass (left panel) and the total transverse energy (right panel).
  Results are shown for two different subtraction schemes, the
  Catani-Seymour (CS) dipole subtraction and the new Nagy-Soper (NS)
  scheme.  The lower panels show the ratio of the results within the
  two schemes.}
\end{figure}
%
The results have also been compared at the differential
level. Differential cross sections in the 5FS as a function of the
total transverse energy, $H_T$, and the invariant mass of the four
bottom  system, $M_{b\bar{b}b\bar{b}}$, are depicted in Figure
\ref{fig:CS-NS}.  Again we observe full agreement between the
predictions calculated with the two schemes.


\section{Summary and Outlook}

We report on the next-to-leading order calculation for  the production
of four bottoms quarks at the LHC at the centre-of-mass energy of
${\sqrt{s} = 14}$ TeV.  The higher order corrections significantly
reduce the scale dependence, with a residual theoretical uncertainty
of about $30\%$ at NLO. The impact of the bottom quark mass is
moderate for the cross section normalization and negligible for the
shape of distributions.   As a completely technical detail,  results
for inclusive and differential cross-sections have been shown for two
subraction schemes  for treating real radiation corrections:
Catani-Seymour and Nagy-Soper.  They provide a validation of our
implementation of the second scheme for
massive and massless fermions within \textsc{Helac-Dipoles}.


\acknowledgments 

This research was supported in part by the German Research Foundation
under Grant No. WO 1900/1-1 (``Signals and Backgrounds Beyond Leading
Order. Phenomenological studies for the LHC'').


\end{document}